\documentclass[runningheads]{llncs}
\usepackage[utf8]{inputenc}
\usepackage{graphicx}
\usepackage{hyperref}
\usepackage{cleveref}
\usepackage{tabularx}
\usepackage{booktabs}
\usepackage{listings}
\usepackage{lmodern}
\usepackage{menukeys}
\usepackage{threeparttable}
\usepackage{geometry}
\usepackage{array}
\usepackage[toc,page]{appendix}

\usepackage[inkscapeformat=pdf]{svg}
\usepackage{url}

\newmenucolortheme{custom}{HTML}{FAFAFA}{555555}{222222}
\changemenucolortheme{menus}{custom}
\begin{document}
\title{What’s Next, Cloud? A Forensic Framework for Analyzing Self-Hosted Cloud Storage Solutions}

\titlerunning{What’s Next, Cloud?}
\authorrunning{Külper et al.}
\author{Michael Külper\inst{1}\and Jan-Niclas Hilgert\inst{1}\and Frank Breitinger\inst{2}\and Martin Lambertz\inst{1}}
\institute{
Fraunhofer FKIE \\
\email{\{michael.kuelper, jan-niclas.hilgert, martin.lambertz\}@fkie.fraunhofer.de} 
\and
University of Augsburg \\
\email{frank.breitinger@uni-a.de}
}

\maketitle

\begin{abstract}
Self-hosted cloud storage platforms like Nextcloud are gaining popularity among individuals and organizations seeking greater control over their data. 
However, this shift introduces new challenges for digital forensic investigations, particularly in systematically analyzing both client and server components. 
Despite Nextcloud’s widespread use, it has received limited attention in forensic research.
In this work, we critically examine existing cloud storage forensic frameworks and highlight their limitations. 
To address the gaps, we propose an extended forensic framework that incorporates device monitoring and leverages cloud APIs for structured, repeatable evidence acquisition.
Using Nextcloud as a case study, we demonstrate how its native APIs can be used to reliably access forensic artifacts, and we introduce an open-source acquisition tool that implements this approach. 
Our framework equips investigators with a more flexible method for analyzing self-hosted cloud storage systems, and offers a foundation for further development in this evolving area of digital forensics.
\end{abstract}


\section{Introduction}
Cloud storage has become a core element of modern computing, shaping how individuals and organizations store, access, and synchronize data across multiple devices. 
Its widespread adoption is driven in part by the integration of cloud services into major user ecosystems, as companies such as Microsoft, Google, and Apple bundle cloud storage with user accounts, effectively lowering the barrier to entry. 
In parallel, dedicated providers like Dropbox and pCloud offer free or low-cost plans with robust features, further expanding access to cloud-based storage solutions. 
As users increasingly access their data across multiple devices, cloud storage has also become a common element in modern forensic investigations~\cite{hargreaves2024dfpulse}.

While commercial cloud services dominate the landscape, the growing popularity of affordable, low-power computing platforms, such as the Raspberry Pi, combined with privacy concerns, has made self-hosted cloud storage solutions increasingly viable. 
Concurrently, regulatory pressures, including the General Data Protection Regulation (GDPR) and the EU-U.S. Data Privacy Framework, are motivating individuals and organizations to seek greater control over their data. 
These developments point to a trend: forensic investigations must increasingly consider not just client-side artifacts from commercial services, but also server-side data from privately hosted environments.

A range of self-hosted cloud storage platforms has emerged in response to this demand, including ownCloud, Nextcloud, OpenCloud, Seafile, and Pydio. 
While some have evolved into comprehensive groupware platforms with features like calendaring and communication tools, many continue to focus primarily on file storage and synchronization. 
This paper centers on the latter category. 
Although extended features such as calendars, address books, and communication tools (e.g., video, audio, and chat) are forensically significant, the core functionalities of file synchronization and sharing are common to many platforms and present a wide range of unique forensic considerations. 
These include the handling of timestamps, the attribution of user actions to digital artifacts, and the underlying reasons for a file's presence on a given device, each of which warrants focused examination.

Research into the forensic investigation of cloud storage applications is well established.
Over a decade ago, Martini~and~Choo~\cite{martini_12} proposed general guidelines for the forensic analysis of cloud platforms. 
To validate their framework, they conducted a case study on the cloud storage application ownCloud~\cite{martini_13}, defining categories of evidential data and artifact sources, and identifying relevant forensic artifacts. 
Building on this foundation, Teing~et~al.~\cite{teing_19} extended the model to offer a more balanced view---addressing the prior emphasis on client-side analysis---and presented a case study on the Seafile application.

In this paper, we revisit and re-evaluate Teing~et~al.~\cite{teing_19}'s forensic framework in the context of Nextcloud.
We chose Nextcloud for three reasons: 
First, it is among the most widely used self-hosted cloud platforms, yet to date, no comprehensive forensic analysis has been published. 
Second, as a fork of ownCloud, Nextcloud enables direct comparison with the original case study by Martini~and~Choo~\cite{martini_12}, allowing us to examine the evolution of artifacts over more than a decade. 
Third, it allows for evaluating whether cloud storage forensic frameworks remain relevant and identifying needed updates for modern platforms.

To this end, our contributions include:
\begin{itemize}
\item a comprehensive forensic analysis of the Nextcloud client and server applications;
\item an exploration of under- and unexplored aspects relevant to modern forensic investigations in cloud storage environments;
\item a review of existing cloud storage forensic frameworks, accompanied by a revised framework that addresses identified gaps;
\item the release of an open-source forensic tool that facilitates API-based data acquisition from Nextcloud environments.
\end{itemize}

\section{Related Work}
Martini~and~Choo~\cite{martini_12} proposed an early framework for conducting cloud forensic investigations, building on the four phases of the forensic process as defined by NIST: Collection, Examination, Analysis, and Reporting. 
Their model comprises the following phases: (i) Evidence Source Identification and Preservation, (ii) Collection, (iii) Examination and Analysis, and (iv) Reporting and Presentation. 
The framework is iterative, allowing newly discovered evidence sources to initiate a fresh cycle of investigation. 
To demonstrate their model, they conducted a forensic analysis of ownCloud~\cite{martini_13}, where artifact identification was closely tied to the specific functions and features of the application.

Quick~et~al.~\cite{quick14} extended this model by adding a preliminary phase, Commence and Preparation, to define the scope of the investigation, select appropriate tools, and coordinate necessary expertise. They also highlighted challenges such as server volatility and the risk of evidence tampering or deletion by suspects using unknown devices.

Building on these foundations, Teing~et~al.~\cite{teing_19} introduced a seven-phase framework tailored to self-hosted cloud storage environments. 
This model refines the original four phases and introduces three new ones: Readiness (pre-investigation planning and understanding of the environment), Case Investigation Preparation (legal coordination, stakeholder engagement, and technical setup), and Investigation Review and Lessons Learned (post-investigation analysis for future improvement). 
Unlike the original framework, preservation is treated as a distinct step during the Collection and Preservation of Evidence phase. 
Their approach also underscores the importance of aligning evidence collection with the architectural and functional specifics of the cloud application under investigation.

Based on the frameworks, both Martini~and~Choo~\cite{martini_13} and Teing~et~al.~\cite{teing_19} analyzed the cloud storage applications ownCloud and Seafile to identify forensic artifacts.
Apart from studies on applications that support self-hosting and allow for server-side analysis, there is also a range of research focused solely on the client side of cloud storage services hosted by cloud storage service providers.
This includes works on Dropbox~\cite{quick_13_2,martini_15,mehreen_15,daryabar_16,lim_20}, OneDrive (formerly SkyDrive)~\cite{quick_13,martini_15,daryabar_16,easwaramoorthy_16}, GoogleDrive~\cite{quick_14,daryabar_16}, Box~\cite{martini_15,daryabar_16}, iCloud~\cite{oestreicher_14}, Amazon Drive~\cite{hale_13,easwaramoorthy_16}, CloudMe~\cite{teing_18}, pCloud~\cite{dargahi_17,mohtasebi_17,ahmad_20}, IDrive~\cite{thamburasa_16,rochmadi_20}, MEGA~\cite{thamburasa_16,daryabar_17,mishra_22}, Sync.com~\cite{bhat_19}, FlipDrive~\cite{bhat_19}, SpiderOak~\cite{mohtasebi_17}, and JustCloud~\cite{mohtasebi_17}.

\section{Forensic Analysis of Nextcloud}
This section presents the results of our forensic analysis of the Nextcloud ecosystem.
Unlike previous approaches based on system architecture, our findings are organized by forensic relevance---focusing on user behavior and actionable evidence.
We identified seven key artifact categories: User Data and Accounts, Device Information, Files, Synchronization, User Activity and File Version, File Deletion and Trash, and File Sharing.
The following subsections summarize the client and server analysis.

\subsection{Experimental Setup}
To identify forensic artifacts in Nextcloud, we analyzed both the Android client and a personal server instance.

Our client-side analysis focused on the official Nextcloud Android app (v3.31.1), chosen for its popularity (over 1M downloads) and open-source availability\footnote{\url{https://github.com/nextcloud/android}}. 
The app was executed on a rooted virtual device via Android Studio, enabling full access to internal storage and system-level activity. 
Forensic artifacts were primarily found in the internal SQLite database \texttt{filelist} (\texttt{/data/user/0/com.nextcloud.client/databases}), which contains 13 tables tracking file metadata, directories, and user interactions. 

On the server side, we examined a long-running NextcloudPi instance (v23.0.0.10) used continuously for over four years. 
It supports a single user syncing across devices, mostly using default settings except for a relocated data directory. 
Forensic data was collected from the live system, but backup ZIP archives containing full SQL dumps (\texttt{nextcloud-sqlbkp\_<date>.bak}, with 101 tables) are also available. 
While our analysis used the live instance, these .bak files provide a viable option for offline inspection---though they may not include the most recent data.
Key configurations reside in \texttt{/var/www/nextcloud/config/config.php}, including database credentials, paths, and logging options such as \texttt{trashbin\_retention\_obligation}. 
Additionally, \texttt{nextcloud.log} offers a valuable source of system event records.

While our analysis focused on application-level artifacts, OS-level and web server logs were excluded. 
To validate the observed behaviors, we performed controlled experiments complemented by source code reviews.

\subsection{User Data and Accounts}
This section outlines how user data and account information are stored and managed on both the client and server sides of a Nextcloud installation.

\subsubsection{Client-side}
The file \texttt{com.nextcloud.client\_preferences.xml}, located in the \texttt{shared\_prefs} directory of the internal storage, contained the app's account information.
For self-hosted instances, the \texttt{select\_oc\_account} field stores an identifier in the format \texttt{<username>@<host>[:<port>]}.

Further metadata is stored in the internal \texttt{filelist} database. 
Initially, the \texttt{arbitrary\_data} table may contain only a path to media folders. 
However, after login, multiple entries for users identified by the \texttt{cloud\_id} value are added, including avatar information, user status, and other metadata.

The \texttt{capabilities} table stores server-related metadata linked to each account. 
While the database includes entries for all configured accounts, the XML file retains only the currently active one.

\paragraph{Stored Credentials and Authentication Artifacts}
\label{sec:credentials}

After login, the client receives an \texttt{appPassword} saved by Android’s Account Manager in the \texttt{accounts\_ce.db} database.
Each Nextcloud account creates entries in the \texttt{accounts} and \texttt{auth\_tokens} tables that store this password.
It is used with the username for HTTP Basic Authentication in a base64-encoded header.

These credentials are significant for forensic analysis, enabling access to the same Nextcloud instance and linking network activity to specific devices. Additional metadata such as the server base URL, version, and user display name is found in the \texttt{extras} table, with separate entries for each account.

Additional account-related metadata, such as the server’s base URL, version information, and the user’s display name, is stored in the \texttt{extras} table of the \texttt{accounts\_ce.db} database. In cases where multiple Nextcloud accounts are configured on the device, corresponding entries are maintained for each of them.

\subsubsection{Server-side}
Nextcloud's database includes several tables that store user-related information. A summary of these tables, along with the relevant forensic data they contain, is provided in \cref{tab:userrelatedtables}.

\begin{table}[htbp]
\centering
\footnotesize
\caption{Overview of Nextcloud User-Related Database Tables}
\label{tab:userrelatedtables}
\begin{tabular}{p{2.5cm} p{11.4cm}}
\toprule
\textbf{Table} & \textbf{Description} \\
\midrule
\texttt{oc\_users} &
Contains a unique identifier (UID) for each user, referenced across multiple tables. Includes display name and password. Passwords are stored using a hashed format: \texttt{algorithm|hash}, generated by PHP’s \texttt{password\_hash()} function. \\

\texttt{oc\_user\_status} &
Tracks user availability status (online, offline, away), along with optional custom messages and the timestamp of the last status change. \\

\texttt{oc\_accounts\_data} &
Stores user profile information set during account creation. Includes structured fields like \texttt{displayname}, \texttt{address}, \texttt{phone}, and \texttt{twitter}. \\

\texttt{oc\_accounts} &
A simplified table containing only \texttt{uid} and \texttt{data}. The \texttt{data} column contains the same profile information in JSON format. \\

\texttt{oc\_groups}, \texttt{oc\_group\_user} &
Define and manage group memberships. Used to distinguish roles such as \texttt{admin} and \texttt{user}. \\

\texttt{oc\_preferences} &
Stores per-user application settings. Includes \texttt{userid}, \texttt{appid}, \texttt{configkey}, and \texttt{configvalue}. For instance, \texttt{core|lang} stores UI language; \texttt{login|lastLogin} stores the last login timestamp. \\

\texttt{oc\_authtoken} &
Stores authentication tokens for users. Upon login via client applications, two rows are created: one for the client and one for the browser session. Each row includes UID, \texttt{last\_activity}, and user agent string. \\
\bottomrule
\end{tabular}
\label{tab:nextcloud-user-tables}
\end{table}

\subsection{Device Information}
This section describes the types of device-specific information that can be identified to support the attribution of user activity to specific devices.

\paragraph{Client-side}
From the client-side, dedicated artifacts for identifying the device are generally redundant, as the analysis is already being conducted on the device itself. 
However, the application password, which is uniquely generated for each client during the authentication process, can serve as an implicit device identifier. 
This token enables the correlation of client activity with server-side logs and observed network traffic, effectively linking actions to a specific device instance. 
Note, however, that if a user logs out and logs back in, a new application password is issued, which prevents consistent device identification across sessions.

\paragraph{Server-side}
The web interface under \menu[,]{Settings,Personal,Security} displays connected devices, including current and past sessions.
This includes user agent strings that reveal device details such as phone models (e.g., Samsung SM-S926B), operating systems (e.g., Android, macOS), browsers (e.g., Firefox 137), and sync clients (e.g., DAVx5 or macOS dataaccessd).
The same information can be retrieved via the database table \texttt{oc\_authtoken} table by running \texttt{SELECT id, uid, name,  last\_activity FROM oc\_authtoken;}.
We observed that each time a new OAuth token is generated, a new row is added to this table.
Additionally, the \texttt{nextcloud.log} file---which location is specified in \texttt{config.conf}---contains entries related to failed login attempts, providing further visibility into authentication activities.

\subsection{Files}
In this section, we describe where to find user data and metadata stored by Nextcloud.

\subsubsection{Client-side}
On Android, Nextcloud maintains metadata for files and directories in a local SQLite database \texttt{filelist}. Each file or directory is represented by a record in the table \texttt{filelist}, which contains the following file metadata:

\begin{description}
  \item[Hierarchy] Name, relative path to root directory, and a parent ID referencing the immediate parent. The root directory is represented as a special entry.
  \item[Content] MIME type (e.g., \texttt{DIR}, \texttt{text/plain}) and file size. File contents themselves are not stored.
  \item[Metadata] Timestamps for creation and modification (typically only \textit{modified} is set, while \textit{created} timestamp was often missing or set to zero). Includes permissions, an owner account string, and—if applicable—image dimensions in \texttt{metadata\_size}.
  \item[Attributes] Flags indicating whether the item is marked as favorite, hidden, or currently downloading.
  \item[Identification] Each entry includes a \texttt{local\_id}, \texttt{remote\_id}, and a primary key.
\end{description}

The database is populated incrementally as the user navigates the file structure. 
As a result, it may reflect only a partial view of the remote file hierarchy at any given time. Deleted files may persist until the user revisits their parent directory, triggering a refresh. 
Consequently, any interaction with the app after acquisition can modify or remove potentially valuable artifacts. 
Forensic practitioners should ensure the database is backed up before conducting further analysis.

In multi-account setups, all file and directory metadata is stored in the same \texttt{filelist} database. 
Entries are distinguished using the \texttt{file\_owner} field, which contains the corresponding account identifier.

Importantly, our experiments indicate that deleting an account from the Nextcloud application does not immediately remove its associated metadata from the local database.
Instead, existing records persist and are gradually overwritten as the app continues to operate. 
This behavior allows for the recovery of remnants related to previously configured accounts during forensic analysis.

\subsubsection{Server-side}
By default, Nextcloud stores user data in \texttt{/var/www/nextcloud/data/}, but the actual location is set by the \texttt{datadirectory} key in the configuration file. 
In our setup, this points to a different folder, referred to as \texttt{\$DATADIR}. 
Each user has a subfolder named after their UID. 
\texttt{\$DATADIR} also includes log files (\texttt{nextcloud.log}, \texttt{updater.log}), a \texttt{tmp/} directory for temporary data, and two system folders\footnote{\url{https://help.nextcloud.com/t/how-to-clean-up-nextcloud-data-appdata-and-updater/123746}}: \texttt{updater-*}, which stores pre-update backups for rollback, and \texttt{appdata-*}, which holds app-specific data. 

Nextcloud also supports external storage\footnote{\url{https://docs.nextcloud.com/server/23/admin_manual/configuration_files/external_storage_configuration_gui.html}}, which can refer to remote sources (SMB, FTP, S3, other Nextclouds) or local paths/drives. 
These are configured via the web GUI (\menu[,]{Administration,External Storage}) or the \texttt{oc\_external\_config} table (\cref{tab:external_config}). 
The type (e.g., SMB, FTP) can be found in \texttt{oc\_external\_mounts}.

The \texttt{oc\_filecache} table accelerates file access by caching metadata. 
It tracks filenames, paths, timestamps, sizes, and mimetypes\footnote{Mimetypes are mapped via \texttt{oc\_mimetypes}; directories typically have \texttt{mimetype = 2}}. Useful queries include:
\begin{itemize}
    \item Recently modified files (within the last day, i.e., 86400 seconds): \\
    {\small\texttt{SELECT name, mtime FROM oc\_filecache WHERE mtime > (UNIX\_TIMESTAMP() - 86400);}}
    \item Files matching \texttt{report.pdf}: \\
    {\small \texttt{SELECT fileid, path, size FROM oc\_filecache WHERE name LIKE '\%report.pdf\%';}}
    \item Top 10 largest files: \\
    {\small\texttt{SELECT name, size FROM oc\_filecache WHERE mimetype != 2 ORDER BY size DESC LIMIT 10;}}
\end{itemize}

\subsection{Synchronization}
To keep the client up to date with changes on the server, Nextcloud relies on a process called synchronization. 
For Nextcloud, this process is triggered by the client.
This section covers what kind of synchronization metadata can be identified within artifacts.

\subsubsection{Client-side}
Each entry in the \texttt{filelist} table within the \texttt{filelist} database contains the following information regarding synchronization:
\begin{itemize}
    \item \texttt{last\_sync\_date}: A millisecond Unix timestamp indicating the last time a file or directory was synced. For root directories of a share, this value seems to be set and stay at \texttt{0}.
    \item \texttt{last\_sync\_date\_for\_data}: A millisecond Unix timestamp indicating the last time the actual content of a file or directory was synced.
    This also applies to the root directory of a share.
    \item \textit{ETag:} Instead of relying on differences in modification times, which can be unreliable in certain circumstances, Nextcloud uses ETags to track versions of files. These randomly chosen numbers are updated when a file changes and are thus used to signal the need for synchronization. 
    \item \texttt{media\_path}: This field is populated when a file is downloaded to the device and specifies the local file system path where the file is stored.
\end{itemize}

Manually refreshing the file view, such as by pulling down the screen, updates synchronization-related timestamps within the \texttt{filelist} database. 
Similarly, navigating through files and directories in the application also causes the synchronization timestamps of the corresponding entries to be updated, even if no file is downloaded or modified.

\paragraph{Downloads}
Nextcloud allows users to download individual files or entire directories for offline access. In the \texttt{filelist} table, such files can be identified by the presence of a value in the \texttt{media\_path} field, which points to the file’s local storage path on the device. The default download location can be configured within the application’s settings and is stored in the shared preferences XML file under the key \texttt{storage\_path}, for example: \texttt{/storage/emulated/0/Android/media/com.nextcloud.client}.

The application creates a subdirectory named after the account identifier within the storage location. Downloaded content is saved there, maintaining the structure of the shared directory as seen by the user.
When a file is downloaded, its \texttt{last\_sync\_date\_for\_data} timestamp is updated to reflect synchronization with the server.

\paragraph{Uploads}
The Nextcloud application allows users to upload arbitrary files, including photos and videos captured directly from the device’s camera. A record of all uploads performed by the application is maintained in the \texttt{list\_of\_uploads} table within the \texttt{filelist} database. Each entry includes the local source path (\texttt{local\_path}), the remote target path on the server (\texttt{remote\_path}), the file size, and a Unix timestamp (in milliseconds) indicating the completion time of the upload.

While an upload is still in progress, the file size is set to \texttt{-1} and the completion timestamp remains \texttt{NULL}. The associated account is recorded in the \texttt{account\_name} column using the known account identifier format. In our experiments, entries in this table were not automatically removed and remained accessible for several weeks after the upload had completed.

\paragraph{Synchronized Folders}
When selecting a directory, the Android application offers the option to `Sync' it. In our experiments, this action resulted in the corresponding update of records in the \texttt{filelist} table, which belong to any files or subdirectories of the synced directory, similar to the automatic sync that is performed when a directory is opened. However, the sync is performed recursively for all layers of the hierarchy, e.g., when a new file is created or deleted multiple layers down compared to the synced directory. In case no changes are detected, the table is not altered.

If \textit{two-way sync} is enabled for folders, it sets the \texttt{internal\_two\_way\_sync\_timestamp} value for the corresponding record in the \texttt{filelist} table from \texttt{-1} to \texttt{0} for the first time. 
Afterwards, the folder is synced in certain intervals from 15 minutes to 24 hours as defined in the application's settings. 
The status as well as the interval can be found in the shared preferences XML file under the \texttt{two\_way\_sync\_status} and \texttt{two\_way\_sync\_interval} keys, respectively. Each time a synchronization is triggered, the timestamp stored in \texttt{internal\_two\_way\_sync\_timestamp} is updated to reflect the corresponding execution time.

\subsubsection{Server-side}
Since Nextcloud synchronization is primarily initiated and managed by client applications, limited synchronization metadata is directly available on the server side. 
However, information about recently synchronized files and folders can be inferred from database tables such as \texttt{oc\_filecache} and \texttt{oc\_activity}. 
As previously mentioned, the \texttt{oc\_filecache} table indirectly reflects synchronization activity: any modification, upload, or update to a file or folder alters the \texttt{mtime} (modification time) and \texttt{storage\_mtime} (storage modification time) fields. 
Therefore, to identify the ten most recently modified files, the following SQL query can be used:
\texttt{SELECT name, path, mtime FROM oc\_filecache ORDER BY mtime DESC LIMIT 10;}.

\subsection{User Activity and File Versions}
Nextcloud allows users to modify files, tag files, and keep different file versions after changes are made.
Within the following we look what kind of information can be retrieved via the artifacts about this user activities. 

\subsubsection{Client-side}
The Android application does not offer a dedicated interface for managing file versions. However, version-related events---such as the creation of a new version---are visible in the application’s \textit{Activities} view, which provides a summary of recent interactions including file creation, renaming, and modification. 
For version events, the view includes a button to restore a previous version. In our testing, attempts to use this restore function consistently resulted in application crashes.

It is important to note that the activity data shown in this view is not stored locally on the device. Instead, it is retrieved dynamically from the server when the user opens the view. As a result, this information is not accessible for offline forensic analysis or in cases where server communication is no longer possible.

\subsubsection{Server-side}
File versions are accessible via the web portal under \menu[,]{Details,Versions}, or by querying the \texttt{oc\_filecache} table with \texttt{path LIKE '\%files\_versions\%'}. 
On disk, they are stored in the \texttt{datadirectory} under each user's \texttt{files\_versions} directory, preserving the original folder structure and appending a UNIX timestamp to filenames (e.g., \texttt{File.docx.v1746040832}).

User activity is recorded in the \texttt{oc\_activity} table, with key columns including \texttt{timestamp}, \texttt{type}, \texttt{user}, \texttt{affecteduser}, \texttt{subject}, and \texttt{subjectparams}. 
The \texttt{type} column indicates the general action category (e.g., \texttt{file\_changed}), while \texttt{subject} and \texttt{subjectparams} provide more specific context.

We found 12 distinct \texttt{type} values (\texttt{addressbook}, \texttt{calendar}, \texttt{calendar\_event}, \texttt{calendar\_todo}, \texttt{file\_created}, \texttt{group\_settings}, \texttt{shared}, \texttt{card}, \texttt{file\_changed}, \texttt{file\_deleted}, \texttt{file\_restored}, and \texttt{security}) and 34 unique \texttt{subject} entries. 
The distinction between \texttt{user} and \texttt{affecteduser} allows us to differentiate who performed an action and who was impacted by it. For example, if \texttt{userA} deletes their own file, the \texttt{subject} includes the suffix \texttt{\_self}; if they delete a file belonging to \texttt{userB}, this suffix is absent.

File version and activity log retention is managed by the \menu[,]{versions\_retention\_obligation,auto} setting, which triggers automatic cleanup based on internal policies.

\subsection{File Deletion and Trash}
Similar to file versioning, Nextcloud retains deleted files in a trash bin for a defined period. 

\subsubsection{Client-side}
Deleting a file using the Android application merely causes its entry to be removed from the \texttt{filelist} table. While the Nextcloud application allows users to browse through deleted files and also recover them, this information is not kept anywhere on the device itself. Furthermore, a downloaded and deleted file is immediately removed from the storage location on the device.

\subsubsection{Server-side} 
When a file is deleted through the system interface, an entry in the \texttt{oc\_activity} database table with the type \texttt{file\_deleted}  is created. 
The file is then moved to the trash directory located at \texttt{datadirectory/username/files\_trashbin/files}. 
Similar to file versioning, deleted files are appended with a timestamp suffix (e.g., \texttt{screenshot.jpg.d1728237675}) to distinguish them from other entries.
Metadata regarding the original file path is retained in the \texttt{oc\_files\_trash} table. 
Notably, if a file is manually removed from disk (e.g., via the \texttt{rm} command), its corresponding entry may persist in \texttt{oc\_files\_trash} until a cleanup operation is triggered.

\subsection{File Sharing}
A core feature of cloud storage applications, is the ability to share files with other users or groups. 

\subsubsection{Client-side}
Nextcloud allows users to share files and directories with other users as well as using a link. 
When a sharing link is created, an entry in the \texttt{ocshares} table in the \texttt{filelist} database is created. Most importantly, this entry contains the path of the shared file or directory, a Unix timestamp indicating the time the share was created, the share URL as well as the user who shared it. 
When sharing with a user, the entry contains the user's name in the \texttt{shate\_with} [sic] column. 
For a shared file, the \texttt{share\_by\_link} attribute in the \texttt{filelist} table is set to \texttt{1}.

\subsubsection{Server-side}
The web portal provides a tab that lists all shares, i.e., files shared with the user as well as files shared by a user. The same information can be found in the table \texttt{oc\_share}, where the most relevant columns are 
\texttt{share\_type} (in our testing, we only encountered the values 3 for link-sharing and 0 for sharing directly with another user),
\texttt{shared\_with}, which is NULL for link-sharing or the corresponding user otherwise, and
\texttt{stime}, which is the timestamp when the share was created. Other information that can be found in the table is note (to recipient), password (if protected), the file name, expiration (if set), or the token, which is how link-shared files are accessed (e.g.,  \nolinkurl{https://pi-ip/index.php/s/3kXXKCtn7WyyNS3} the token is 3kXXKCtn7WyyNS3).

\subsection{Artifact Comparison: Nextcloud vs. ownCloud}
The comparison with ownCloud is based on the server side on ``Cloud storage forensics: ownCloud as a case study''~\cite{martini_13} and on the client side on both this and ``Mobile Cloud Forensics''~\cite{martini_15} as both contain information about Android ownCloud artifacts.

On the server side, we were able to verify that most of the findings identified for ownCloud still apply to Nextcloud. 
One of the main differences, however, lies in the file sharing mechanism. 
Specifically, the \texttt{oc\_sharing} table used in ownCloud has been replaced by a table called \texttt{oc\_share} in Nextcloud.
While the \texttt{oc\_sharing} table in ownCloud contained only basic information—such as the file owner, the users the file was shared with, and their access rights (read-only or write access)---the \texttt{oc\_share} table in Nextcloud includes more detailed metadata. 
As previously described, this includes the type of share (e.g., with another user or via a public link), an optional password for protected shares, and an expiration date for time-limited access.
Additionally, we identified several aspects that were not covered in the analysis of ownCloud, either because they did not exist at the time or were not examined. 
These include new database tables and changes in how the application handles file deletions.

Similar to the server side, many of the existing artifacts identified in ownCloud are still present in Nextcloud. 
Those that have changed appear to be associated with specific feature updates.
One key difference lies in how passwords are handled.
While both applications utilize the Android Account Manager API, the analysis revealed that ownCloud stores the username and password in plaintext. 
In contrast, Nextcloud stores a revocable \texttt{appPassword}, which is used for authentication instead.
Moreover, the Nextcloud client introduces two additional tables in the \texttt{filelist} database (\texttt{arbitrary\_data} and \texttt{capabilities}), which store user-specific and server instance–specific information, respectively.

Summarized we can see that a lot of the identified artifacts also exist for ownCloud in the same or a very similar way, as they are related to certain features that got reworked throughout update in features such as file sharing or authentication. 
It is unclear whether these changes may also be found in an investigation of an modern ownCloud instance or if their artifacts may differ.

\section{Rethinking Cloud Storage Forensic Frameworks}
Beyond the selected artifact classes, we identified additional aspects that existing research and frameworks have overlooked. 
This section highlights these underexplored areas and introduces an enhanced forensic framework for the analysis of cloud storage applications.

\subsection{Underexplored Aspects}

\paragraph{API Acquisition}
A common feature of cloud storage applications is an API provided by the storage server, typically used by client or web applications to perform operations such as uploading, downloading, or managing data. 
In a forensic context, this API can offer a practical entry point for accessing server-side data, particularly in early investigative stages when the server's physical location is not yet known or accessible. 
Even in self-hosted scenarios, where the server might eventually be identified and examinable, the API provides an opportunity for timely and remote data acquisition.

Despite the ubiquity of such APIs, their forensic potential has received limited attention in existing research, even though they can provide access to server-side artifacts that may not be present on the client device. 
One of the few comprehensive studies exploring this area is the work by Roussev~et~al.~\cite{roussev16}, which evaluates services like Google Drive, Microsoft OneDrive, Dropbox, and Box, and emphasizes the benefits of API-based acquisition in forensic investigations.

When considering API-based collection, it is important to evaluate the possible impact on the server’s state. 
Interactions through the API may alter or generate server-side artifacts, which can compromise the forensic soundness of subsequent analyses. 
This is especially relevant in self-hosted environments, where the server might later become available for physical examination. 
Some API requests may also trigger logging, alerts, or other forms of administrative attention. 
Therefore, investigators should adopt a cautious and controlled approach, favoring selective and granular data acquisition over large-scale extraction, to reduce detection risks and preserve the integrity of the investigation.


\paragraph{Server Evidence Preservation Process}
Another often overlooked aspect is the careful consideration required when preserving evidence from a live server instance.
While a dead acquisition can eliminate the risk of user tampering or unintentional corruption of files and databases, the benefits of live forensics often outweigh this, particularly because it enables the retrieval of volatile information such as cryptographic keys and potentially active session data.

To reduce the risk of users modifying or destroying evidence during live acquisition, investigators may deauthenticate active users and block new logins.
However, this can alert users and prompt evidence tampering.
A more discreet approach is to display a generic maintenance message while disabling functionality, or to isolate the server from the network.
While these methods help prevent interference, their likelihood of raising suspicion varies by context and relies on the investigator’s judgment.

\paragraph{Monitoring}
Maintaining system operation beyond an initial preservation and collection phase is crucial for capturing ongoing activities and uncovering additional forensic insights. 
Continuous monitoring provides significant value, particularly for server devices, where observation can reveal critical events such as the creation of new files, user logins from previously unknown devices, or rapid data deletions. 
These actions may highlight files of interest or indicate usage patterns that suggest targeted behavior directed at specific users or datasets.
On the client side, monitoring is also vital. 
It can help identify when shared files with the inspected user are deleted, when access permissions are revoked, or when new data is shared. 
Furthermore, API-based monitoring is beneficial for retrieving data and metadata that may not be persistent on the client device. 
Repetitive checks for changes via the server can offer valuable insights into potential alterations, similar to server-based monitoring.

\paragraph{File Attribution}
File attribution involves linking files to specific users, groups, or conceptual user entities based on actions such as uploading, creating, editing, deleting, sharing, or accessing. 
Attribution metadata varies across artifact types, and accurately tracing user actions and ownership requires careful collection and interpretation of this data.

However, attribution can be challenging. 
For instance, anonymous file drops obscure the uploader’s identity. 
Similarly, federated sharing---where interconnected application instances enable cross-instance file access---can complicate attribution depending on the implementation.
A comprehensive attribution analysis would require executing a broad range of user actions and systematically monitoring resulting artifact changes. 
As this necessitates an automated framework, it lies beyond the scope of this work and is left for future research.



\paragraph{Further Aspects}
A further challenge arises in multi-server architectures where databases or cloud storage reside on external servers. This setup complicates data acquisition, often requiring API access. Although API usage may be feasible in certain cases, it typically demands authentication credentials configured within the server—details which may be accessible during forensic analysis. However, due to its rarity and dependency on backend-specific implementations, this approach is not pursued in the present work.

Cloud storage applications also differ in their internal file storage strategies. Some use file fragmentation for deduplication, necessitating chunk identification and reassembly prior to analysis. Others apply delta encoding, storing only modified file segments across versions.

In network traffic analysis, techniques such as chunked uploads and differential syncing—where only file segments are transferred—can complicate reconstruction.

Since Nextcloud does not implement fragmentation, delta encoding, chunked uploads, or differential syncing, these mechanisms fall outside the scope of this study. Likewise, features specific to individual applications or client software, such as multi-account support or virtual files, are not addressed.

\subsection{An improved Cloud Storage Forensic Framework}
Building upon the underexplored aspects identified in the previous section, we propose a refined forensic framework tailored to the challenges of modern cloud storage environments.

A key distinction of the newly proposed framework is the inclusion of a monitoring phase. 
As discussed in the previous section, we believe that continuous monitoring can significantly enhance the forensic value of investigations, providing deeper insights into data access and usage patterns.
To streamline the model, we exclude three phases from Teing~et~al.'s~\cite{teing_19} framework: Readiness, Case Investigation Preparation, and Investigation Review and Lessons Learned. 
These are general to forensic practice and not specific to cloud storage forensics.

\begin{figure}
    \centering
    \includegraphics[width=0.47\linewidth]{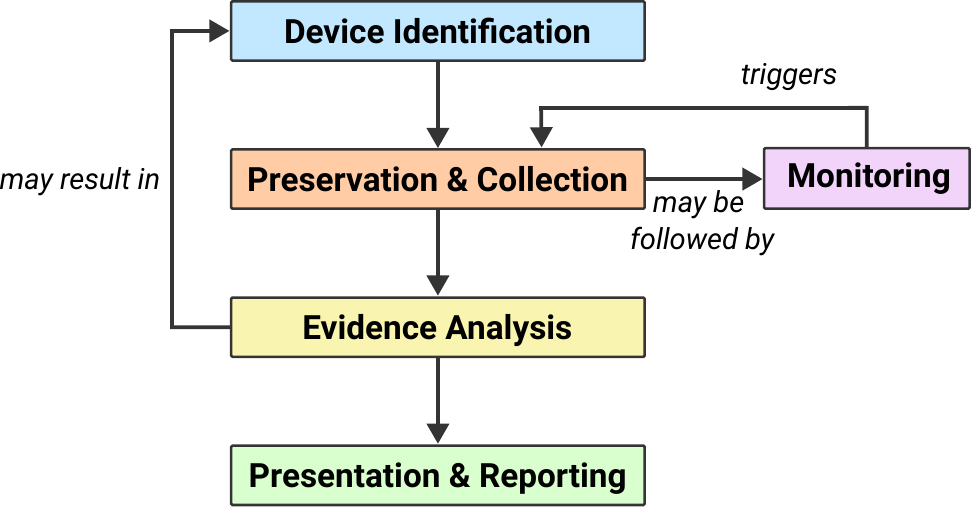}
    \caption{An improved cloud storage forensic framework including a new Monitoring phase}
    \label{fig:framework}
\end{figure}

Other identified gaps do not require structural modifications but instead call for refinements within individual phases.
Additionally, some gaps pertain to specific features or scenarios and are therefore not addressed in the overall framework overview.
Consequently, our proposed framework consists of five phases, as illustrated in \Cref{fig:framework} and detailed below:
\begin{enumerate}
    \item \textit{Device Identification}: The first phase involves identifying devices such as mobile phones, laptops, computers, or servers. 
    Device identification can occur by physically locating the device with a suspect or by identifying unique identifiers, such as a server's IP address or associated domain.
    \item \textit{Preservation \& Collection}: In this second phase, the first objective is to preserve the data by creating a copy, ideally unmodified, from the data source. 
    This is often achieved through classic acquisition methods like drive imaging and memory dumping, but can also be achieved through API acquisition.
    The collection aspect focuses on extracting relevant evidence, typically through the retrieval of artifact content. 
    Depending on the data source and the tools used, preservation and collection may be performed simultaneously, as exemplified for API analysis.
    We therefore decided to place both Preservation \& Collection (as already done by Teing~et~al.~\cite{teing_19}) into a single phase.
    \item \textit{Monitoring}: The monitoring phase involves continuous observation of a device. 
    Based on detected changes or after a set time interval, the preservation and collection of selected or all data sources may be repeated. 
    This phase effectively becomes an iterative extension of the Preservation \& Collection phase. 
    While this ongoing process is generally advisable, it may not always be feasible in practice due to legal limitations or time constraints.
    \item \textit{Evidence Analysis}: Following the Preservation \& Collection phase, evidence analysis involves scrutinizing the extracted artifacts to identify and extract relevant data for the investigation.
    Common categories of artifacts include user data and accounts, device information, files and file metadata, synchronization metadata, user activity, file versions, file deletion and trash, and file sharing. 
    Additionally, further devices, such as cloud storage server instances or other devices with client instances, may be identified. 
    Depending on the case, these additional devices may undergo similar analysis, starting with the Preservation \& Collection phase, thus triggering an iteration within the framework.
    \item \textit{Presentation \& Reporting}: This phase is identical to the phase of the same name described by Teing~et~al.~\cite{teing_19}, emphasizing the structured communication of findings through clear documentation and, when necessary, expert testimony. It ensures that evidence is conveyed accurately and comprehensibly to stakeholders, including legal professionals, while maintaining the integrity and admissibility of the data.
\end{enumerate}

\section{API Acquisition}
Building on our earlier discussion of API access, this sections examines the server-side artifacts retrievable through API-based acquisition for Nextcloud.

\subsection{API Analysis}
In this section, we present our findings on relevant Nextcloud API endpoints, as shown in \cref{tab:nextcloud-api-endpoints}, which can be leveraged for forensic acquisition.
We also evaluate the forensic soundness of these endpoints to assess their suitability for use in evidence collection.

We identified two primary APIs that can be used to query information from a Nextcloud instance: the Open Collaboration Services (OCS) API\footnote{\url{https://docs.nextcloud.com/server/latest/developer_manual/client_apis/OCS/index.html}} and the WebDAV API.\footnote{\url{https://docs.nextcloud.com/server/latest/developer_manual/client_apis/WebDAV/index.html}} 
As outlined in Section~\ref{sec:credentials}, both APIs support HTTP Basic Authentication, which can be performed using the account username and an application-specific password obtained from a previously acquired client device.
Alternatively, if the user’s primary password is available, a new app password can be generated.

\subsubsection{User Data and Accounts}
Nextcloud provides the OCS API endpoint \texttt{ocs/v2.php/cloud/user} to query information about the currently authenticated user, i.e. , the user that belongs to the provided app password.  This endpoint returns metadata such as the user's display name, email address, storage quota, and other account-related details. 

Additionally, the \texttt{ocs/v1.php/cloud/users} endpoint allows administrators to retrieve a list of all users on the server. To obtain more detailed information, such as last login time or group memberships, about each user, the \texttt{ocs/v2.php/cloud/users/{username}} endpoint can be used. While administrators can query information for all users using this endpoint, default users can only obtain information about themselves. Another endpoint is \texttt{ocs/v2.php/cloud/users/details}, which requires a \texttt{search} parameter to search for user details based on a search term.

\subsubsection{Device Information}
Nextcloud’s public APIs do not provide direct access to device metadata. 
However, device and session details are available via the web interface under \textit{Devices \& Sessions} (\texttt{settings/user/security}). 
Each entry typically represents a device linked to an application-specific password and includes metadata such as session name (e.g., browser or app identifier), device type (e.g., mobile or desktop), and the timestamp of last activity.

Users with appropriate privileges can manage individual sessions through this interface: sessions can be revoked (invalidating the associated application password), file system access can be restricted, and a remote wipe can be initiated. The remote wipe command both revokes server access and instructs the client to delete locally stored Nextcloud data.

\subsubsection{Files}
Nextcloud exposes a WebDAV interface at \texttt{remote.php/dav/files/{username}} for enumerating user files and directories via the HTTP \texttt{PROPFIND} method. The request includes an XML body specifying desired properties, and the response returns metadata such as IDs, names, paths, sizes, MIME types, timestamps, and ownership. The \texttt{Depth} header (commonly set to \texttt{1}) controls directory traversal depth.

Metadata for specific files or directories can be queried by appending their full path to the endpoint. The response includes a \texttt{<d:href>} element indicating the resource location, which can be used in a subsequent HTTP \texttt{GET} request to retrieve the content.
For property specifications and request structure, see the official documentation\footnote{\url{https://docs.nextcloud.com/server/20/developer_manual/client_apis/WebDAV/basic.html}}.

\subsubsection{Synchronization}
The WebDAV API endpoint \texttt{remote.php/dav/files/} provides metadata such as \texttt{etag} and \texttt{getlastmodified}, which can be used to infer file and directory changes and potential synchronization activity. 
However, it does not expose explicit synchronization events or client-specific timestamps. 
Such data is typically maintained on the client side, as demonstrated earlier for the Android application, limiting the server API’s utility for reconstructing precise synchronization timelines.

\subsubsection{User Activity and File Versions}
For retrieving a file's activity history the OCS API provides the endpoint \texttt{ocs/v2.php/apps/activity/api/v2/activity/filter}. 
By specifying the file’s \texttt{object\_id} as a GET parameter, this endpoint returns a list of recorded file activities. Each entry includes an activity ID, a type (e.g., creation, modification, or deletion), and a corresponding timestamp. Notably, activity entries remain accessible through this endpoint even after the associated file has been permanently deleted from the trash bin.

For versioning, Nextcloud offers the \texttt{remote.php/dav/versions/{username}/versions/{fileId}} endpoint, which returns a list of all available versions of a given file. 
Each version entry contains a last modified timestamp, an \texttt{etag}, and a reference path, where the timestamp serves as a unique identifier for the version (e.g., \texttt{remote.php/dav/versions/{username}/versions/{fileId}/{timestamp}}).

This path can be used to retrieve metadata or download the file content corresponding to that specific version.

\subsubsection{Deleted Files and Trash}
A users's trash bin is accessible via a WebDAV \texttt{PROPFIND} request to the endpoint \texttt{remote.php/dav/trashbin/{username}/trash}. 
This returns deleted files and directories with metadata such as display name, size, and unique identifier. 
Additional trashbin-specific properties include: \texttt{trashbin-location} (name in the trash), \texttt{trashbin-original-location} (original path), and \texttt{trashbin-deletion-time} (Unix timestamp).

Restoring a file involves a \texttt{MOVE} request to \texttt{remote.php/dav/trashbin/{username}/restore/}, which reintegrates it into the user’s file hierarchy. 
As this operation alters the cloud environment, we instead locate and directly download deleted files from the trash to preserve the existing storage structure for forensic purposes.

\subsubsection{File Sharing}
The OCS API endpoint \texttt{/ocs/v2.php/apps/files\_sharing/api/v1/shares} can be used to query the shares of a user supporting several optional parameters, such as \texttt{reshares}, \texttt{shared\_with\_me}, and \texttt{subfiles}, which allow filtering the results to include reshares, shares received by the user, or subfiles within shared directories, respectively. The response contains detailed metadata about each share, including the share type (e.g., user, group, or public link), permissions, expiration date, and the path to the shared resource.

\subsection{Implementation}
To support the acquisition and forensic analysis of Nextcloud environments, we developed and publicly released a set of specialized tools\footnote{\url{https://github.com/miqsoft/nextcloudforensic}}. 
While general-purpose tools like rclone---a command-line program supporting various cloud storage backends---exist, they are not designed for forensic purposes and lack support for metadata extraction (e.g., activity logs, file versions, deletion traces)~\cite{breitinger2022forensic}.
Our tool focus on preserving exactly these artifacts to ensure forensic soundness and transparency.

The implementation supports data retrieval and examination via Nextcloud's APIs and web interface. 
It typically requires the instance URL and user credentials (username and application-specific password).

Since Nextcloud does not expose a public API for device/session management, we reimplemented relevant web client functionality. 
Our approach directly interacts with specific web endpoints to retrieve and revoke connected devices while avoiding full web interface emulation, thereby minimizing the risk of unintended state changes.

Our implementation offers the following capabilities:

\begin{itemize}
\item \textbf{User Information:} Retrieve user-related data, such as detailed account info, user listings, and user search capabilities (\texttt{user-info}, \texttt{list-users}, \texttt{search-user}).
\item \textbf{File System Analysis:} List files, resolve file IDs to paths, and download file contents \\ (\texttt{list-files}, \texttt{file-id-to-path}, \texttt{download-file}).
\item \textbf{Deleted Content Recovery:} Access and extract information from the trash bin \\ (\texttt{trash-bin}).
\item \textbf{File History and Activity:} Retrieve file version history and activity logs \\ (\texttt{file-versions}, \texttt{file-activity}).
\item \textbf{Device Management:} List connected devices and revoke their access \\(\texttt{list-devices}, \texttt{revoke-device}, \texttt{revoke-all}). This allows investigators to prevent further interaction with the cloud storage from other devices during forensic acquisition.
\item \textbf{Full Share Dump:} Recursively acquire the entire content of a user’s file share (\texttt{dump}).
\end{itemize}

Detailed usage instructions and documentation are available in the project’s repository.

\subsubsection{Nextcloud: The Sleuth Kit (TSK)-Inspired Utilities}
To provide a familiar interface for practitioners, we developed a set of command-line tools inspired by The Sleuth Kit, adapted to cloud-based storage:

\begin{tabularx}{0.96\textwidth}{@{} l X @{}}
\textbf{\texttt{nc-fsstat}} & Displays general information about the Nextcloud instance, including server capabilities and available users. \\
\textbf{\texttt{nc-fls}} & Provides a file system view similar to traditional forensic tools. Supports specifying a subdirectory, recursive traversal (\texttt{-r}), and showing deleted files (\texttt{-d}) marked with an asterisk. Deleted files are identified by correlating trash bin data with the current hierarchy. Each file or directory is listed with its object ID. \\
\textbf{\texttt{nc-istat}} & Retrieves metadata about a file or directory based on its object ID, including file activity and available versions. \\
\textbf{\texttt{nc-icat}} & Outputs the content of a file identified by its object ID. \\
\end{tabularx}

\section{Conclusion}
Cloud storage systems challenge traditional forensic methods due to their distributed, dynamic nature and limited client-side visibility. 
This work shows how API-based acquisition offers a practical, forensically aware entry point for accessing server-resident data---especially in early investigation stages. 
When combined with continuous monitoring, API integration enables timely, targeted capture of both static and ephemeral evidence, enhancing visibility into user actions and system changes across client and server environments.

Centered on Nextcloud, our analysis demonstrates how native APIs, often overlooked in previous research, can provide structured, reliable, and repeatable access to critical artifacts. 
To support this, we developed and released an open-source acquisition tool tailored to Nextcloud’s API, but extendable to further cloud storage applications.

Our analysis revealed that many artifacts---persisting even after more than a decade---remain consistent with those identified in earlier studies of ownCloud, the software from which Nextcloud was forked.
Apart from the artifacts, we identified a major shortcoming in existing cloud storage forensic frameworks: their reliance on static, one-time evidence collection of a device.
This approach misses opportunities to detect events after a first acquisition, such as new file uploads, user interactions, or newly connected client devices---all of which may be crucial during an investigation. 
In response, we proposed a revised forensic framework incorporating a monitoring phase and refining several stages to better reflect the demands of modern, self-hosted environments.

Our work lays the groundwork for further research in this area. While it analyzes existing artifacts, it does not comprehensively map artifact changes to specific user actions---such as creating, sharing, or deleting files—as this would require an automated system capable of simulating a wide range of user scenarios.
Many modern self-hosted platforms also include groupware features like calendars, messaging, and video conferencing, which may contain valuable forensic artifacts. 
Exploring whether the API acquisition method and our framework can be extended to these services is a promising avenue for future work.

While we have conceptually outlined the adapted forensic framework for analyzing self-hosted cloud storage applications, validating it through real-world case studies remains a key task for future research.
Finally, while decentralized systems like SyncThing differ significantly from cloud-based platforms, the iterative, monitoring-focused design of our framework may still be applicable—offering another opportunity for adaptation and evaluation.

\bibliographystyle{splncs04}
\bibliography{bibliography.bib}

\appendix
\clearpage

\section{Server Side Database Tables}
\begin{table}[htbp]
\caption{Content of the server side database \texttt{oc\_external\_config}}
\label{tab:external_config}
\centering
\begin{threeparttable}
\begin{tabular}{llll}
\toprule
\textbf{config\_id} & \textbf{mount\_id} & \textbf{key} & \textbf{value} \\
\midrule
1 & 1 & datadir  & /media/crypt/Cloud/folder1 \\
2 & 2 & datadir  & /home/hans/folder2 \\
5 & 3 & datadir  & /var/foobar/folder3 \\
19 & 6 & host     & thisIs.MyURL.de \\
20 & 6 & root     &  \\
21 & 6 & secure   & 1 \\
22 & 6 & user     & ftpSecrect-user \\
23 & 6 & password & 75779a1b1fd4105186fda1250a70f\textbar d51ce64f0bfbb167dfc... \\
\bottomrule
\end{tabular}
\end{threeparttable}
\end{table}

\section{Nextcloud API Endpoints}

\setlength{\tabcolsep}{5pt} 
\begin{table}[htbp]
\centering
\caption{Key Nextcloud API Endpoints for Forensic Analysis by Information Type}
\label{tab:nextcloud-api-endpoints}
\begin{tabularx}{\textwidth}{l X l}
\toprule
\textbf{Information Type} & \textbf{API Endpoint(s)} & \textbf{Method} \\
\midrule
User Data &
\texttt{ocs/v2.php/cloud/user} \newline
\texttt{ocs/v1.php/cloud/users} \newline
\texttt{\detokenize{ocs/v2.php/cloud/users/{username}}} &
GET \\
\midrule
Server Capabilities &
\texttt{ocs/v1.php/cloud/capabilities} &
GET \\
\midrule
File System Data &
\texttt{remote.php/dav/files/\{username\}} \newline
\texttt{remote.php/dav/files/\{username\}/\{path\}} &
PROPFIND \\
\midrule
Deleted Files &
\texttt{remote.php/dav/trashbin/\{username\}/trash} &
PROPFIND \\
\midrule
File Versions &
\texttt{remote.php/dav/versions/\{username\}/versions/\{fileId\}} \newline
\texttt{remote.php/dav/versions/\{username\}/versions/\{fileId\}/\{timestamp\}} &
PROPFIND, GET \\
\midrule
Activity Data &
\texttt{ocs/v2.php/apps/activity/api/v2/activity/filter} &
GET \\
\midrule
Sharing Data &
\texttt{ocs/v2.php/apps/files\_sharing/api/v1/shares} &
GET \\
\midrule
Security \& Sessions &
\texttt{ocs/v2.php/core/apppassword} \newline
\texttt{ocs/v2.php/core/apptokens} &
DELETE, GET \\
\bottomrule
\end{tabularx}
\end{table}

\end{document}